\documentclass[usenatbib]{mn2e}
\usepackage{graphicx, amssymb, aas_macros, natbib,array}
\usepackage{mathrsfs}
\usepackage{amsmath}
\usepackage{gensymb}
\usepackage{mathrsfs}
\usepackage{txfonts}

\newcommand{\vrot}{v_{\rm{rot}}}
\newcommand{\mvir}{M_{\rm{vir}}}
\newcommand{\mhalo}{M_{\rm{halo}}}

\newcommand{\rvir}{R_{\rm{vir}}}

\newcommand{\mstar}{M_{\rm \star}}
\newcommand{\msun}{\rm \, M_{\odot}}

\newcommand{\hmpc}{h^{-1} \, \rm Mpc}

\newcommand{\kpc}{\rm \, kpc}
\newcommand{\pc}{\rm \, pc}
\newcommand{\gyr}{\rm \, Gyr}

\newcommand{\kms}{{\rm km \, s}^{-1}}

\newcommand{\lcdm}{$\Lambda$CDM}

\newcommand{\ufdtwo}{\rm UFD \, 2}
\newcommand{\ratio}{\vrot / \sigma}

\newcommand{\ra}[1]{\renewcommand{\arraystretch}{#1}}

\begin{document}

\title[Rotation vs dispersion support in dwarfs]
{The no-spin zone: rotation vs dispersion support in observed and simulated dwarf galaxies
}\author[C. Wheeler et al.]{Coral Wheeler$^1$\thanks{$\!$crwheele@uci.edu},
Andrew B. Pace$^{1,2}$, James S. Bullock$^{1}$, Michael Boylan-Kolchin$^{3}$
\newauthor Jose O\~{n}orbe$^{4}$, Oliver D. Elbert$^{1}$, Alex Fitts$^{3}$, Philip F. Hopkins$^{5}$, Du\v{s}an Kere\v{s}$^{6}$\\ 
\noindent$\!\!$ $^{1}$Center for Cosmology, Department of Physics and Astronomy,
University of California, Irvine, CA 92697, USA \\
\noindent$\!\!$ $^{2}$Department of Physics and Astronomy, Mitchell Institute for Fundamental Physics and Astronomy, Texas A\&M University, College Station, \\
TX 77843-4242, USA \\
\noindent$\!\!$ $^{3}$Department of Astronomy, The University of Texas at Austin, 2515 Speedway, Stop C1400, Austin, TX 78712\\
\noindent$\!\!$ $^{4}$Max-Planck-Institut fuer Astronomie, Koenigstuhl 17, 69117 Heidelberg, Germany\\
    \noindent$\!\!$ $^{5}$TAPIR, Mailcode 350-17, California Institute of Technology, Pasadena, CA 91125, USA\\
    \noindent$\!\!$ $^{6}$Department of Physics, Center for Astrophysics and Space Sciences, University of California at San Diego, 9500 Gilman\\
    Drive, La Jolla, CA 92093}

 \pagerange{\pageref{firstpage}--\pageref{lastpage}} 
 \pubyear{2015}
\maketitle
\label{firstpage} 

\begin{abstract} 
We perform a systematic Bayesian analysis of rotation vs. dispersion support ($\ratio$) in 40 dwarf galaxies throughout the Local Volume (LV) over a stellar mass range $10^{3.5} \msun < \mstar < 10^{8} \msun$. We find that the stars in $\sim 80\%$ of the LV dwarf galaxies studied -- both satellites and isolated systems -- are dispersion-supported. In particular, we show that $6/10$ {\em isolated} dwarfs in our sample have $\ratio \lesssim 1.0$, while all have $\ratio \lesssim 2.0$. These results challenge the traditional view that the stars in gas-rich dwarf irregulars (dIrrs) are distributed in cold, rotationally-supported stellar disks, while gas-poor dwarf spheroidals (dSphs) are kinematically distinct in having dispersion-supported stars. We see no clear trend between $\ratio$ and distance to the closest $\rm L_{\star}$ galaxy, nor between $\ratio$ and $\mstar$ within our mass range. We apply the same Bayesian analysis to four \textsc{FIRE} hydrodynamic zoom-in simulations of isolated dwarf galaxies ($10^9 \msun < \mvir < 10^{10} \msun$)  and show that the simulated {\em isolated} dIrr galaxies have stellar ellipticities and stellar $\ratio$ ratios that are consistent with the observed population of dIrrs \textit{and} dSphs without the need to subject these dwarfs to any external perturbations or tidal forces. We posit that most dwarf galaxies form as puffy, dispersion-dominated systems, rather than cold, angular momentum-supported disks. If this is the case, then transforming a dIrr into a dSph may require little more than removing its gas.
\end{abstract}


\section{Introduction}
\label{sec:intro} 

Dwarf spheroidal (dSph) galaxies comprise the largest population of galaxies in the Local Group, consisting of nearly $60$ confirmed members \citep{Kleyna2005, Munoz2006, Martin2007, Simon2007, Simon2011, Simon2015, Aden2009, Belokurov2009, Carlin2009, Geha2009, Koch2009, Walker2009, Walker2015b, Kalirai2010, Collins2010, Collins2011, Koposov2011, Koposov2015, Koposov2015b, Willman2011, Tollerud2012, Collins2013, Kirby2013, Kirby2015, Kirby2015b, Tollerud2013, Laevens2015a, Laevens2015, Kim2015, Kim2015b, Martin2015b, Martin2015}. These objects are characterized by their low luminosities, spheroidal shapes, high mass-to-light ratios, and by the absence of appreciable gas or recent star formation \citep{Ferguson1994, vandenBergh1999, Mateo1998, Dalcanton2004, Yoachim2006, McConnachie2012}. Line of sight velocity measurements suggest that dSphs have little to no rotation in their stellar populations and velocity dispersion profiles that are nearly flat with radius \citep{Wilkinson2004, Munoz2005, Munoz2006b, Walker2006, Walker2007, Koch2007a, Koch2007b, Mateo2008}.

In the Local Group, dSphs tend to occupy regions close to either the Milky Way or M31 \citep{Mateo1998, Grebel1999}. At greater distances from the two massive galaxies, the population of dSphs dwindles and gives way to a different class of low-mass galaxies called dwarf irregulars (dIrrs). These galaxies have similar luminosities to dSphs, but are distinct most notably in that they have retained some of their gas. Many dIrrs also demonstrate disky features and rotation in their H{\sc I} content \citep{Mateo1998, McConnachie2012}. This ``Local Group morphology-density relation," with dSphs found close and dIrrs found far from MW and M31, mimics similar relationships between galaxy shape and distance from the local barycenter found in clusters \citep{Oemler:1974qf, Dressler:1980pd}. This, and the fact that both dSphs and dIrrs can be fit with exponential light profiles \citep{Mateo1998, Ferguson1994, Faber1983}, is often used to argue in favor of a dwarf irregular transformation-based origin for dSphs \citep{Faber1983, Mayer2001a}. If, as is commonly understood from classical galaxy formation theory, all galaxies initially form as thin, angular momentum supported disks \citep{White1978, Fall1980, Blumenthal1984}, then significant transformation must occur to convert these rotationally-supported galaxies into the puffy, dispersion-dominated dSphs we see today.

The currently-favored mechanism for bringing about this transformation is known as ``tidal stirring" \citep{Mayer2001b, Mayer2001a}. According to this model, rotationally-supported dwarfs with exponential stellar disks and high gas fractions are repeatedly tidally shocked at the pericenters of their orbits. While ram pressure is primarily responsible for removing gas from the dwarf, it is the repeated tidal shocks that produce the morphological transformation. In general, for low-mass dwarfs (the majority of those found in the Local Group), this involves the creation of a tidally-induced bar, which transports high angular momentum material to the outer regions of the galaxy where it is subsequently stripped. This reduces the rotation of the system and transforms the galaxy into a spheroidal, dispersion-supported system \citep[][and references therein]{Mayer2010}. In the tidal stirring model, a galaxy is generally considered to have been transformed into a dSph if it has no (or very little) gas, an ellipticity within a specific range (usually $0.1 < e < 0.5$; greater values of ellipticity indicate a more elongated shape), and if the ratio of its line-of-sight rotational velocity to its velocity dispersion, $\ratio$, is below some value -- usually $1$, but as low as $0.5$.  A number of early simulations investigating this effect had considered infalling dIrr models with extremely cold disks ($\ratio \simeq 5$) but more recent simulations involve somewhat hotter initial disks~\footnote{ \citet{Kazantzidis2013} suggest that if dark matter halos are more core-like, then it would be natural to consider $\ratio \simeq 1 - 1.5$ as starting points because $v_{\rm rot}$ is reduced at small radii while $\sigma$ might be expected to stay fixed.} $\ratio \simeq 3$ \citep[e.g.][]{Kazantzidis2011a}. 

While tidal stirring simulations have been successful at producing systems with $\ratio \lesssim 1$ \citep{Mayer2001b, Mayer2001a, Mayer2006, Klimentowski2009, Mayer2010, Lokas2011, Kazantzidis2011a, Kazantzidis2013, Lokas2015, Tomozeiu2015},  historically it has been difficult to reduce $\ratio$ to values $< 0.5$ found for many observed dwarf satellites \citep{Mastropietro2005}. 
The most complete transformations occur for highly eccentric orbits \citep{Mayer2001b, Mayer2001a}, at low inclination, and that are mildly prograde \citep[but see \citealt{Mayer2006}]{Kazantzidis2011a, Lokas2015}. The high eccentricity in particular allows for shorter orbital times and repeated pericenter passages (typically $3-5$, but as many as $8$). Short orbital times ($1-3~\gyr$) and close pericenter distances ($10-70~\kpc$) have been shown to be particularly important to the transformation \citep{Kazantzidis2011a}. Interestingly, these simulations have often found that the accreted dIrr galaxies need to orbit within a Milky Way host potential for $\sim 10~\gyr$ in order to be able to complete the required number of pericenter passages \citep{ Klimentowski2009, Mayer2010, Lokas2011, Kazantzidis2011a, Kazantzidis2013, Lokas2015,Tomozeiu2015}.

One major issue with any scenario that requires $\sim 10$ Gyr in order to transform a dIrr to a dSph is that this is quite long compared to the expected accretion times for satellites derived from cosmological simulations of Milky Way and Local Group analogues.  Specifically, the overwhelming majority of Milky Way satellites are predicted to have fallen in less than $10~\gyr$ ago \citep{Rocha2012, Garrison-Kimmel2014a, Fillingham2015, Wetzel2015}, with $\sim 40\%$ accreted within the last $4$ Gyr. Only $2$ of the $11$ classical Milky Way dwarf satellites are dIrrs \citep[which appear to have fallen in very recently,][]{Besla2007}, and they are significantly more massive than the dSph satellites. This suggests that any environmental transformation associated with dSph formation needs to occur within $\sim 2$ Gyr of accretion \citep{Fillingham2015,Wetzel2015}.  Furthermore, at least two dSphs, Cetus and Tucana, currently exist at large distances from either the Milky Way or Andromeda \citep[$681$ and $882~\kpc$ from the closest giant,][]{McConnachie2012}. They, like the dSphs much closer to their hosts, have little to no gas and their stars are dispersion rather than rotationally-supported (see below). Explaining the existence of such distant objects as the result of tidal stirring poses a particularly difficult challenge to the model. Due to this difficulty, \citet{Kazantzidis2011a} predict that distant dSphs should have systematically higher values of $\ratio$. Alternatively, it has been shown that dwarfs with highly cored dark-matter profiles undergo faster transformations \citep[after just $1-2$ pericenter passages,][]{Kazantzidis2013, Tomozeiu2015}. This reduction in required time spent near the host would be particularly useful in explaining the lack of rotation in an object like Leo I, which has undergone only a single pericenter passage at a distance of $\sim 100~\kpc$ from the Milky Way \citep{Sohn2013, Boylan-Kolchin2013}. 

There are other alternative mechanisms for transforming a dIrr into a dSph that require the initial galaxy to interact with another object. Dwarf-dwarf mergers can create dSphs \citep{Moore:1996kl,Kazantzidis2011b}, and the mechanism is satisfyingly similar to models proposed for transforming massive disks into giant ellipticals \citep{Icke1985}. \citet{Starkenburg2015} propose that the spheroidal shapes of dSphs can be reproduced by mergers between dwarf galaxies and lower-mass dark halos, but do not discuss rotation support. Another model, ``resonant stripping", posits that a fly-by between a dwarf and a galaxy 100 times its mass can instigate resonances in the smaller dwarf that preferentially strip the stellar material \citep{DOnghia2009}. Interactions between dwarfs in the Local Group are common \citep{Deason2014}, but merger-based transformation scenarios fail to explain the ``Local Group morphology-density relation," and so are not likely to account for a large fraction of observed dSphs.

Given the strict requirements for the tidal stirring mechanism to be effective, it seems reasonable to question the initial conditions used for dwarf galaxies in these models. The traditional picture of disk galaxy formation was developed for massive galaxies \citep{Fall1980,Blumenthal1984} with virial temperatures $T_{\rm v} \sim 10^6~K$, which is well above the expected bulk ISM temperature of a cooled gas in a galaxy $T_{\rm g} \sim 10^4$ K.  In this case, the pressure support radius of cooled gas will be tiny compared to the angular-momentum support radius.~\footnote{The radius of pressure support declines exponentially as the ratio $T_{\rm g}/T_{\rm v}$ shrinks, where $T_{\rm g}$ is a phenomenological proxy that mimics the net effect of velocity dispersion from various feedback effects, such as inefficient cooling, heating by an internal or external ultraviolet (UV) background, supernova feedback, turbulent pressure, or cosmic-ray heating, among others \citep{Kaufmann2007}.} It is in this sense that the disk of a massive galaxy is expected to be ``cold."   However,  \citet[][hereafter KWB]{Kaufmann2007} show, using a simple analytic approximation and hydrodynamic simulations, that low-mass galaxies with shallow potential wells and modest virial temperatures ($T_{\rm v} \lesssim 10 ~T_{\rm g}$) will tend to have pressure support radii that are comparable to their angular-momentum support radii. KWB did not look at $\ratio$ explicitly, but showed that at low virial mass ($M_{\rm vir} \lesssim 10^{11} \msun$), the dispersion-supported component of a galaxy should begin to rival the rotationally-supported component \citep[see also e.g.][]{Dalcanton2010}. As first suggested by \citet{Read2006}, the above arguments only strengthen if one considers additional ISM pressure imparted on small galaxies from internal feedback effects and turbulent motions. Moreover, stars, unlike gas, can never re-cool after their orbits are disturbed by potential fluctuations or mergers. Taken together, these arguments suggest that the stellar populations of dwarf galaxies residing in the field are not necessarily expected to exhibit well-ordered, disk-like motions as seen in their larger cousins.  

Recently, large samples of stellar kinematic data for local dIrr galaxies have become available \citep{Simon2007, Fraternali2009,  Leaman2009, Leaman2012, Kirby2014}. These data enable more detailed studies of the pressure support in field dwarfs. In particular, \citet{Kirby2014} present a stellar kinematic analysis of seven (non-satellite) dwarf galaxies in the local volume, and showed that only one among them (Pegasus) demonstrates a clear sign of rotation in its stellar population. While they did not explicitly rule out rotation in the other objects, the work of Kirby et al. provides some suggestion that a high degree of rotation support is not the rule among isolated dwarfs.

In what follows, we conduct a systematic search for stellar rotation in Local Group dwarfs. We use a Bayesian analysis on a large observational sample of dwarfs consisting of twenty eight MW and M31 dSphs, two dwarf ellipticals (dEs), and ten dwarfs beyond the virial radii of either the MW or M31 (including two isolated dSphs and eight dIrrs) to estimate $\ratio$. We confirm previous findings that both the MW and M31 dSphs, with few exceptions, have stellar populations that are not rotating. We show further that isolated dwarfs in the Local Group are also largely dispersion-supported, with only two of ten showing strong Bayesian evidence for rotation, and seven of ten failing to show even moderate evidence in favor of rotation. We propose an alternative formation scenario for dSphs galaxies: most dwarf galaxies form initially as puffy, dispersion-supported or slowly rotating systems, and gas removal via ram pressure stripping (enabled by internal feedback) is likely the main process that leads to the formation of dSphs. We demonstrate the feasibility of this in a \lcdm~scenario by using the same Bayesian analysis to measure the rotation support in four hydrodynamic cosmological zoom-in simulations of \textit{isolated} dwarf galaxies run with FIRE/\textsc{gizmo}. The star particles in our simulated isolated dwarf galaxies are dispersion-supported, without any interaction with a more massive galaxy, and their ellipticities are also similar to the known dSph population without the need for harassment. 

In Section \ref{sec:obs}, we highlight our observational sample. Our simulated dwarfs and their characteristics are described in Section \ref{sec:sims}. Section \ref{sec:method} is used to explain the Bayesian analysis we perform on each galaxy. The results of our systematic search for stellar rotation are given in Section \ref{sec:res}. We discuss these findings in Section \ref{sec:dis} and conclude in Section \ref{sec:summ}.

\begin{table*}
	\ra{1.3}
	\centering 
	\tabcolsep=0.11cm
	\begin{tabular}{| l   c  c  c  c | c   c  c  c   c c  c |} 
		\hline
		$\rm Dwarf$ &  $\rm Category$ & $\mstar (10^6 \msun)$ & $\rm d_{L_\star}  (\kpc)$ & $\rm ellipticity$  & $\ratio$ &  $\vrot \, (\kms) $ & $\sigma \, (\kms)$ & $\rm N$ & $\ln{B_{\rm rad}}$ &  $\ln{B_{\rm rot}}$ & Ref.\\
		\hline
		\hline
		Coma Berenecis  &  UF dSph  & $0.0037$   &  $45$  & $0.38_{-0.14}^{+0.14}$ & $1.78_{-0.92}^{+1.07}$ & $7.01_{-3.24}^{+3.16}$ & $4.00_{-0.87}^{+0.95}$  & $59$ & $3.13$ & $0.73$ & (a), (b) \\

		Ursa Major II & UF dSph  & $0.0041$ & $38$  & $0.63_{-0.05}^{+0.05}$ & $0.81_{-1.41}^{+1.07}$ & $5.33_{-9.27}^{+6.04}$ & $6.46_{-1.43}^{+1.88}$  & $20$ & $1.57$ & $-0.17$ & (a), (b) \\

		Canis Venatici II & UF dSph  & 0.0079    &  161  & $0.52_{-0.11}^{+0.11}$ & $1.35_{-0.57}^{+0.67}$ & $5.21_{-1.91}^{+1.83}$ & $3.88_{-0.96}^{+1.19}$  & $25$ & $4.21$ & $1.64$ & (a), (b) \\
	
		Ursa Major I & UF dSph  & $0.014$ & $102$ & $0.80_{-0.04}^{+0.04}$ & $0.15_{-0.41}^{+0.41}$ & $1.20_{-3.23}^{+3.25}$ & $7.93_{-1.01}^{+1.25}$  & $39$ & $1.05$ & $-1.73$ & (a), (b) \\

		Bootes I & UF dSph  & 0.029   &    64  & $0.39_{-0.06}^{+0.06}$ & $0.28_{-0.25}^{+0.28}$ & $1.44_{-1.30}^{+1.42}$ & $5.16_{-0.48}^{+0.55}$  & $74$ & $1.19$ & $-2.00$& (a), (c) \\
		
		Hercules & UF dSph  & 0.037   &    126 & $0.68_{-0.08}^{+0.08}$ & $0.20_{-0.69}^{+0.59}$ & $1.10_{-3.76}^{+3.25}$ & $5.49_{-0.92}^{+1.11}$  & $30$ & $1.38$ & $-1.76$ &  (a), (c) \\

		Canis Venatici I & UF dSph  & 0.23   &     218  & $0.39_{-0.03}^{+0.03}$ & $0.04_{-0.21}^{+0.18}$ & $0.29_{-1.61}^{+1.38}$ & $7.67_{-0.45}^{+0.49}$  & $214$ & $1.38$ & $-2.48$ &  (a), (c) \\

		\hline
		
		Draco & MW  dSph & 0.29    &    76 &$0.31_{-0.02}^{+0.02}$ & $0.29_{-0.10}^{+0.10}$ & $2.62_{-0.86}^{+0.90}$ & $9.05_{-0.28}^{+0.31}$  & $476$ & $2.68$ & $1.99$ &  (a), (d) \\

		Ursa Minor & MW dSph & 0.29   &     78 & $0.56_{-0.05}^{+0.05}$ & $0.21_{-0.18}^{+0.15}$ & $1.87_{-1.59}^{+1.39}$ & $9.02_{-0.32}^{+0.34}$  & $867$ & $1.62$ & $-0.72$  &  (a), (e) \\
	
		Carina &  MW dSph  & 0.38   &    107 & $0.33_{-0.05}^{+0.05}$ & $0.00_{-0.09}^{+0.09}$ & $0.03_{-0.56}^{+0.56}$ & $6.44_{-0.21}^{+0.22}$  & $758$ & $1.84$ & $-2.60$ & (a), (f) \\
	
		Sextans &  MW dSph  & 0.44  &      89 & $0.35_{-0.05}^{+0.05}$ & $0.08_{-0.11}^{+0.12}$ & $0.54_{-0.80}^{+0.82}$ & $7.10_{-0.27}^{+0.30}$  & $424$ & $1.82$ & $-2.30$ & (a), (f) \\

		Leo II &  MW dSph  & 0.74    &    236 & $0.13_{-0.05}^{+0.05}$ & $0.13_{-0.19}^{+0.18}$ & $0.86_{-1.29}^{+1.24}$ & $6.76_{-0.44}^{+0.49}$  & $164$ & $1.26$ & $-2.17$ & (a), (g) \\

		Sculptor &  MW dSph  & 2.3   &      86 & $0.32_{-0.03}^{+0.03}$ & $0.16_{-0.10}^{+0.12}$ & $1.37_{-0.84}^{+1.03}$ & $8.79_{-0.18}^{+0.18}$  & $1349$ & $3.35$ & $-0.72$ & (a), (f) \\

		Sagittarius* &  MW dSph  & 3.5   &      18 & $0.65_{-0.01}^{+0.01}$ & $0.28_{-1.33}^{+0.89}$ & $2.11_{-10.33}^{+6.78}$ & $7.68_{-1.13}^{+1.48}$  & $180$ & $10.83$ & $80.26$ &  (a), (h) \\

		Leo I &  MW dSph  & 5.5    &     258 & $0.21_{-0.03}^{+0.03}$ & $0.16_{-0.12}^{+0.12}$ & $1.44_{-1.09}^{+1.12}$ & $8.99_{-0.37}^{+0.40}$  & $327$ & $2.16$ & $-1.37$ &  (a), (i) \\

		Fornax &  MW dSph  & 20     &     149 & $0.30_{-0.01}^{+0.01}$ & $0.03_{-0.07}^{+0.05}$ & $0.30_{-0.70}^{+0.57}$ & $10.59_{-0.16}^{+0.17}$  & $2409$ & $1.47$ & $-2.76$ &  (a), (f) \\

		\hline
		
		Andromeda XIV & M31 dSph & 0.02  &      162 & $0.20_{-0.11}^{+0.11}$ & $0.24_{-0.43}^{+0.40}$ & $1.42_{-2.61}^{+2.42}$ & $6.10_{-0.91}^{+1.08}$  & $48$ & $1.20$ & $-1.34$3 &  (a), (j), (k)\\

		Andromeda X & M31 dSph  & 0.096  &     110 & $0.30_{-0.18}^{+0.18}$ & $0.39_{-0.50}^{+0.50}$ & $2.85_{-3.66}^{+3.57}$ & $7.37_{-1.32}^{+1.73}$  & $21$ & $1.35$ & $-0.99$ &  (a), (j), (k) \\

		Andromeda IX & M31 dSph  & 0.15  &      40 & $0.12_{-0.07}^{+0.07}$ & $0.10_{-0.43}^{+0.47}$ & $1.22_{-5.21}^{+5.70}$ & $12.08_{-2.05}^{+2.56}$  & $32$ & $1.16$ & $-1.06$ & (a), (j), (k) \\

		Andromeda V & M31 dSph  & 0.39    &    110 & $0.28_{-0.07}^{+0.07}$ & $0.28_{-0.31}^{+0.32}$ & $3.08_{-3.39}^{+3.52}$ & $11.02_{-1.08}^{+1.28}$  & $85$ & $1.87$ & $-0.70$ & (a), (j), (k) \\

		Andromeda XV & M31 dSph  & 0.49    &    174 & $0.24_{-0.10}^{+0.10}$ & $0.42_{-0.63}^{+0.57}$ & $2.08_{-2.56}^{+2.87}$ & $4.92_{-1.34}^{+1.67}$  & $29$ & $1.65$ & $-1.04$ & (a), (j), (k) \\

		Andromeda III & M31 dSph  & 0.83     &    75 & $0.59_{-0.03}^{+0.03}$ & $0.60_{-0.58}^{+0.57}$ & $5.87_{-5.45}^{+5.57}$ & $9.74_{-1.24}^{+1.58}$  & $62$ & $0.35$ & $-0.35$ & (a), (j), (k) \\

		Andromeda VI & M31 dSph  & 2.8      &   269 & $0.41_{-0.03}^{+0.03}$ & $0.04_{-0.54}^{+0.55}$ & $0.54_{-7.19}^{+7.36}$ & $13.29_{-2.11}^{+2.75}$  & $38$ & $1.10$ & $-0.78$ & (a), (j), (l) \\

		Andromeda I & M31 dSph  & 3.9      &   58 & $0.29_{-0.03}^{+0.03}$ & $0.48_{-0.57}^{+0.68}$ & $5.31_{-6.50}^{+7.60}$ & $11.17_{-1.82}^{+2.36}$  & $51$ & $0.94$ & $-0.31$ & (a), (j), (k) \\

		Cassiopeia III & M31 dSph  & 3.98    &    144 & $0.50_{-0.09}^{+0.09}$ & $0.25_{-0.37}^{+0.24}$ & $2.04_{-3.01}^{+2.00}$ & $8.30_{-0.50}^{+0.54}$  & $212$ & $1.88$ & $-1.48$ & (m), (n) \\

		Lacerta I & M31 dSph  & 6.3    &     275 & $0.43_{-0.07}^{+0.07}$ & $0.13_{-0.28}^{+0.24}$ & $1.36_{-2.91}^{+2.51}$ & $10.30_{-0.74}^{+0.83}$  & $127$ & $1.32$ & $-1.65$ & (m), (n) \\

		Andromeda II & M31 dSph & 7.6    &     184 & $0.14_{-0.02}^{+0.02}$ & $1.43_{-0.17}^{+0.18}$ & $11.43_{-1.33}^{+1.31}$ & $7.97_{-0.37}^{+0.38}$  & $474$ & $6.80$ & $70.13$  & (a), (j), (o) \\

		Andromeda VII & M31 dSph & 9.5    &     218 & $0.13_{-0.04}^{+0.04}$ & $0.47_{-0.20}^{+0.21}$ & $6.11_{-2.59}^{+2.65}$ & $12.95_{-0.97}^{+1.05}$  & $135$ & $2.01$ & $1.25$ & (a), (j), (k) \\

		\hline
		
		NGC 147 & dE/dSph &  62     &     142 & $0.46_{-0.02}^{+0.02}$ & $0.96_{-0.11}^{+0.11}$ & $17.05_{-1.91}^{+1.81}$ & $17.73_{-0.58}^{+0.63}$  & $520$ & $11.42$ & $77.56$ & (a), (j), (p) \\

		NGC 185 & dE/dSph & 187  &   442 & $0.22_{-0.01}^{+0.01}$ & $0.45_{-0.11}^{+0.12}$ & $10.62_{-2.48}^{+2.87}$ & $23.73_{-0.80}^{+0.84}$  & $442$ & $4.13$ & $12.99$ & (a), (j), (p) \\

		\hline
		
		Leo T & Iso dIrr/dSph & 0.14    &    422 & $0.29_{-0.14}^{+0.12}$ & $0.08_{-0.87}^{+1.01}$ & $0.66_{-7.13}^{+8.20}$ & $8.17_{-1.61}^{+2.08}$  & $19$ & $1.80$ & $-0.64$& (a), (b) \\

		Tucana & Iso dSph & 0.56    &    882 & $0.48_{-0.03}^{+0.03}$ & $0.22_{-0.39}^{+0.44}$ & $4.79_{-8.64}^{+8.99}$ & $21.37_{-3.34}^{+4.56}$  & $19$ & $0.71$ & $-0.25$ & (a), (q) \\

		Aquarius $\dagger$ & Iso dIrr/dSph & 1.6    &     1066 & $0.50_{-0.10}^{+0.10}$ & $1.70_{-1.01}^{+1.23}$ & $10.47_{-5.66}^{+5.68}$ & $6.24_{-1.35}^{+1.63}$  & $43$ & $-1.00$ & $0.62$ & (a), (r) \\

		Cetus & Iso dSph & 2.6   &      681 & $0.33_{-0.06}^{+0.06}$ & $0.02_{-0.57}^{+0.34}$ & $0.15_{-4.65}^{+2.75}$ & $8.24_{-0.75}^{+0.84}$  & $120$ & $1.73$ & $-1.46$&  (a), (r) \\

		Leo A & Iso dIrr & 6.0    &     803 & $0.40_{-0.03}^{+0.03}$ & $1.99_{-1.09}^{+0.99}$ & $10.93_{-4.85}^{+5.17}$ & $5.46_{-0.92}^{+1.11}$  & $50$ & $0.29$ & $1.50$ &  (a), (r) \\

		Pegasus & Iso dIrr & 6.61    &    474 & $0.46_{-0.02}^{+0.02}$ & $1.43_{-0.22}^{+0.25}$ & $16.25_{-2.24}^{+2.56}$ & $11.36_{-0.80}^{+0.92}$  & $105$ & $0.19$ & $29.09$ &  (a), (r) \\

		VV 124 & Iso dIrr/dSph & 8.3    &     1367 & $0.44_{-0.04}^{+0.04}$ & $0.56_{-0.57}^{+1.00}$ & $5.22_{-5.25}^{+8.98}$ & $9.27_{-0.93}^{+1.08}$  & $87$ & $1.40$ & $-0.47$ &  (a), (r) \\

		WLM & Iso dIrr & 43      &    836 & $0.65_{-0.01}^{+0.01}$ & $1.01_{-0.15}^{+0.17}$ & $14.79_{-2.04}^{+2.32}$ & $14.69_{-0.81}^{+0.90}$  & $180$ & $10.83$ & $21.36$ & (a), (s), (t) \\

		IC 1613& Iso dIrr & 100  &       517 & $0.24_{-0.06}^{+0.06}$ & $0.48_{-0.63}^{+0.39}$ & $4.99_{-6.57}^{+3.96}$ & $10.44_{-0.71}^{+0.79}$  & $143$ & $3.89$ & $1.46$ & (i), (r) \\

		NGC 6822 $\dagger$ & Iso dIrr & 100    &     452 & $0.24_{-0.05}^{+0.05}$ & $0.41_{-0.15}^{+0.12}$ & $9.38_{-3.46}^{+2.77}$ & $22.62_{-0.92}^{+0.99}$  & $314$ & $-0.26$ & $3.30$ & (a), (r)  \\
		\hline
		
	\end{tabular} 
	\label{tab:tableobs} 
	\caption{Properties and estimated parameters of all galaxies in the observed sample. (1) Name of galaxy. (2) Galaxy type. (3) Galaxy stellar mass from literature. (4) Distance from galaxy to its nearest massive neighbor from literature -- either the Milky Way or M31. (5) Ellipticity of galaxy obtained from literature, with error. (6) Median of parameter $\ratio$ from Bayesian analysis, with $\pm 1~ \sigma$ error. (7) Median rotational velocity from Bayesian analysis, with $\pm 1~ \sigma$ error. (8) Median velocity dispersion from Bayesian analysis, with $\pm 1~ \sigma$ error. (9) Number of stars used in analysis. (10) $\ln{B_{\rm rad}}$, where $B$ is the Bayes factor for the rotation model. Values less than $3$ imply weak/inconclusive evidence for the radially varying model and negative values favor the flat rotation model, see Section \ref{sec:method} for details). (11) $\ln{B_{\rm rot}}$, where $B$ is the Bayes factor for rotation vs non-rotation. Values less than $3$ imply weak/inconclusive evidence for rotation and negative values favor non-rotation to varying degrees, see Section \ref{sec:method} for details). (12) Citations: a) \citealt{McConnachie2012}, b) \citealt{Simon2007} c) \citealt{Koposov2011}, d) \citealt{Walker2015}, e) \citealt{Pace2016}, f) \citealt{Walker2009}, g) \citealt{Koch2007b}, h) \citealt{Frinchaboy2012}, i) \citealt{Mateo1998},  j) \citealt{Salomon2015}, k) \citealt{Tollerud2012}, l) \citealt{Collins2013}, m) \citealt{Martin2013}, n) \citealt{Martin2014}, o) \citealt{Ho2012}, p) \citealt{Geha2010}, q) \citealt{Fraternali2009}, r) \citealt{Kirby2014}, s) \citealt{Leaman2009}, t) \citealt{Leaman2012}. *We exclude Sagittarius from all figures. See Section \ref{sec:indi} for details. $\dagger$ There are only two galaxies for which a flat rotation model is preferred (Aquarius and NGC 6822). For these two galaxies, columns 6,7,8 and 11 are all calculated assuming a flat rotation model.}
\end{table*} 

\section{Observations}
\label{sec:obs}

We analyze spectroscopic data for 40 Local Group galaxies, which are listed by name in Table 1 (column 1) along with the number of stars used in our analysis (column 10).  We use measured line-of-sight velocities for each star as well as the associated errors kindly provided by the authors in the references listed below.   

Among Milky Way satellites, our sample includes all nine of the classical dwarfs: Carina, Fornax, Sculptor, Sextans \citep{Walker2009}, Draco \citep{Walker2015}, Leo I \citep{Mateo2008}, Leo II \citep{Koch2007b}, Sagittarius \citep{Frinchaboy2012}, and Ursa Minor \citep{Pace2016}. For the ultra-faint dSphs of the Milky Way we examine Canes Venatici I, Canes Venatici II, Coma Berenices, Hercules, Leo IV,  Ursa Major I, Ursa Major II \citep{Simon2007}, and Bo\"{o}tes I \citep{Koposov2011}. 

For the M31 system we examine 14 satellites: And II \citep{Ho2012}, And I, And III, And V, And VII, And IX, And X,  And XIII, And XIV, And XV,  \citep{Tollerud2012}, And VI, \citep{Collins2013} Cassiopeia 3, and Lacerta 1 \citep{Martin2014}, NGC 147, and NGC 185 \citep{Geha2010}. 

Finally, we study ten isolated Local Group galaxies: Tucana \citep{Fraternali2009}, Leo T  \citep{Simon2007}, NGC 6822, IC 1613, VV 124, Pegasus dIrr, Leo A, Cetus and Aquarius \citep{Kirby2014}, and WLM \citep{Leaman2009, Leaman2012}.
The dwarf galaxies Phoenix \citep{Irwin2002} and Antlia \citep{Tolstoy2000} have spectroscopic samples that are too small to search for rotation. 

The authors of these spectroscopic studies have taken care to remove foreground contamination. We adopt those same selection criteria here. All samples are homogeneous except for WLM, which consists of data from two distinct observations (one with Keck and the other with the VLT). The analysis includes all stars in each sample, and the samples span varying degrees of spatial extent within the galaxy (the majority go out to $\sim 1.5$ effective radii). All stars analyzed are either red giant or horizontal branch stars.

A subset of our analysis includes an allowance for proper motion (see below). This effect is only important for the satellites of the Milky Way.  We specifically use proper motion measurements from \textit{Hubble Space Telescope} (HST) observations when available.  In the standard frame $(\mu_{\alpha},\mu_{\delta} )$ and in units of $\textrm{mas century}^{-1}$, these are: Carina \citep[$22\pm9, 15\pm9$;][]{Piatek2003}, Draco \citep[$17.7 \pm 6.3,-22.1\pm 6.3$;][]{Pryor2015}, Fornax \citep[$47.6\pm-4.6,-36.0\pm4.1$;][]{Piatek2007}, Leo I \citep[$11.40 \pm 2.95, -12.56 \pm 2.93$;][]{Sohn2013},  Leo II \citep[$10.4\pm11.3,-3.3\pm15.1$;][]{Lepine2011}, Sagittarius \citep[$-254 \pm18, -119 \pm 16$;][]{Massari2013}, Sculptor \citep[$9\pm13, 2\pm13$;][]{Piatek2006}, and Ursa Minor \citep[$-50\pm17,22\pm16$;][]{Piatek2005}.

\section{Simulations}
\label{sec:sims}

Our simulations were previously presented in \citet{Wheeler2015}, and consist of four\footnote{In \citet{Wheeler2015}, we also analyzed two additional simulations that used the same initial conditions as one of our $\sim 10^{10}~\msun$ halos, but were run with slight changes to the subgrid feedback implementation (see \citealt{Wheeler2015} for details). We have not included analysis of those two runs in the text or in the figures here, but note that they have values of $\ratio$ and ellipticity similar to the other runs analyzed here, and so would not change our results if included.} cosmological zoom-in simulations of isolated dwarf galaxy halos. Two were run at the mass of the halos believed to host classical dwarf galaxies ($\mvir \simeq 10^{10} \msun$) and two at lower mass ($\mvir \simeq 10^9 \msun$) (see \citealt{Wheeler2015} for details). All of our simulations were run with the fully conservative cosmological hydrodynamic code \textsc{Gizmo} \citep{Hopkins2014a} in ‘PSPH-mode’, with the standard FIRE feedback implementation. Every run uses a gas particle mass of $m^{\rm gas}_{p}  = 255~\msun$ except for $\ufdtwo$, which uses $m^{\rm gas}_{p} = 499~\msun$. The gas force resolution varies from $\rm \epsilon^{min}_{gas} = 1.0 - 2.8~\pc$, and the stellar masses of the resultant galaxies span $\sim 10^{3.9} - 10^{6.3} \msun$. 

All of these cosmological simulations are of isolated dwarfs, that is, with no large neighbors in either the high or low resolution regions. All but one of the ($\mvir \simeq 10^9 \msun$) dwarfs were selected from $5~\hmpc$ boxes to have typical values of spin parameter $\lambda$, concentration, and formation time for their mass range, and also to have small Lagrangian volumes \citep{Onorbe2014}. The lowest mass ($\mvir \simeq 10^9 \msun$) dwarf was selected from a $25~\hmpc$ box and required to have no other halos of $50\%$ or more of its mass within $4~\rvir$ at $z = 0$ and a small Lagrangian volume. All analysis was performed on the $z=0$ snapshot of each simulation.

\section{Measuring Rotation}

\subsection{Bayesian Analysis}
\label{sec:method}

For each galaxy, we investigate models with and without rotation in order to determine if there is evidence in favor of rotation. We do not assume the stellar components necessarily exist within coherently rotating disks -- the rotation we measure is based entirely on the observed gradient in velocity across the face of the galaxy in the sky. We assume that the likelihood of observing a distribution $\mathcal{D} = (\mathbf{v}, \mathbf{\epsilon})$ of $N$ stars with line-of-sight velocities $v_j$ and associated errors $\epsilon_j$  is:

\begin{equation}
\mathscr{L} = \prod_{j=1}^{N}   \frac{1}{\sqrt{2 \pi (\sigma^2+\epsilon_j^2)}} \exp{\left[-\frac{1}{2} \frac{(v_j - v^{\mathrm{rel}}_j)^2}{\sigma^2+\epsilon_j^2}\right]},
\end{equation}
where $\sigma$ is the underlying (constant) velocity dispersion and $v^{\mathrm{rel}}_j$ is a relative velocity, the form of which depends on whether the model is rotating or non-rotating.
In the absence of rotation, the relative velocity is simply the average bulk motion of the system $v^{\mathrm{rel}}_j = \overline{v}$.  With rotation, the relative velocity becomes
\begin{equation}
\label{vrel}
v^{\mathrm{rel}}_j = \overline{v}  +  \vrot \cos{(\theta - \theta_j)},
\end{equation}
\noindent where $\theta$ is a model parameter (measured from North to East) that defines the axis of rotation, $\theta_j$ is the position angle for each star, and $\vrot$ is the observed rotation across this axis. We explore two models for $\vrot$: constant rotation, $\vrot(R) = v_o$ and a radially varying pseudo-isothermal sphere, $\vrot = v_o \sqrt{1 - \rm R_o/R \arctan(R/R_o)}$, where R is the distance from the rotation axis on the plane of the sky, and $v_o$ and $R_o$ are the rotation velocity and rotation radial scale parameters respectively. With our rotationally varying model, the rotation curve continues to rise to a maximum value. Rotation at radii larger than that reached by the spectroscopic sample is unconstrained, so we choose to measure rotation at $90\%$ of the extent of the data. This choice further prohibits the signal from being dominated by outliers in the sample. We have checked that this radius encloses over $70\%$ of the total mass in each galaxy for over half of the sample (and over $60\%$ of the total mass for $\gtrsim 60\%$ of the sample). Using the radius that encloses $75\%$ or $95\%$ does not significantly change the results presented here.

Note that if the galaxy's angular momentum vector is inclined relative to us with an angle $i$, then  $\vrot = v^{{\rm intrinsic}}_{{\rm rot}} \sin{i}$, where $v^{{\rm intrinsic}}_{{\rm rot}}$ is the magnitude of the intrinsic rotation.  In what follows we quote results for $\vrot$ (rather than $v^{{\rm intrinsic}}_{{\rm rot}}$) because $\sin{i}$ is poorly constrained for the stars. The value of $\vrot$ is a lower limit on the intrinsic value of $v^{{\rm intrinsic}}_{{\rm rot}}$. We discuss the possible effects of inclination in Section \ref{sec:threeD}.

For nearby dwarfs, the line-of-sight velocities as measured from Earth will not project along parallel directions.  One implication is that if a galaxy is moving in the transverse direction, a significant component of this proper motion can be observed as a gradient in the line-of-sight motions of stars across the face of the galaxy \citep{Feast1961, van_der_Marel2002}.  
This perspective proper motion effect can be important for interpreting the dynamics of local galaxies \citep{Kaplinghat2008,Walker2008}
and we therefore include it when possible here.
All classical dSphs except Sextans have proper motion measurements from Hubble Space Telescope (HST) observations.  For
these galaxies, we include the perspective proper motion effects on the relative velocity as
$v_{\mathrm{rel}} \rightarrow v_{\mathrm{rel}} + v_{\mathrm{perspec}}$, marginalizing over the proper motion using 
Gaussian priors centered on the reported measurements (see below). We do not include the (currently unmeasured) proper motion parameters in Sextans or any of the Ultra Faint dSphs. The isolated and the M31 systems are too distant for proper motions to have a measurable effect.

The posterior distribution, $\mathcal{P}(\mathscr{M}| \mathcal{D}, H)$, is the distribution of model parameters $\mathscr{M}$ given the observation of data $\mathcal{D}$.  The symbol $H$ represents the model under consideration: we consider both rotating and non-rotating scenarios.  The likelihood, $\mathscr{L}=\mathcal{P}(\mathcal{D} | \mathscr{M},H)$,  is the probability to observe the data given a set of model parameters.  The posterior is related to the likelihood via Bayes' Theorem:

\begin{equation}
\mathcal{P}(\mathscr{M}| \mathcal{D}, H) = \frac{\mathcal{P}(\mathcal{D} | \mathscr{M},H) Pr(\mathscr{M}) }{ \mathcal{P}(\mathcal{D}, H)},
\end{equation}

\noindent where $Pr(\mathscr{M})$ is the prior distribution, set by our preconceived knowledge of the model.  In our fiducial case that explores rotation and allows for proper motion, we have model parameters $\mathscr{M} = (\overline{v}, \sigma, v_o, \theta, R_o, \mu_{\alpha}, \mu_{\delta})$, where $\mu_{\alpha}$ and  $\mu_{\delta}$ are the proper motions.

The denominator in Equation 3, $Z= \mathcal{P}(\mathcal{D}, H)$,  is referred to as the Bayesian evidence.  It is a normalization factor that is commonly ignored, but will be used for model comparison in our analysis. To test whether the radially varying rotation model is favored over the flat rotation model, we compute the natural log of the Bayes factor, which is defined as the ratio of the evidence for each model: $\ln{B_{\rm rad}}=\ln({Z_{\rm rot,rad} / Z_{\rm rot,flat})}$. A value greater than zero favors the radially varying model. Then, for the preferred model, we compute the natural log of the Bayes factor for the rotating model compared to a model with no rotation: $\ln{B_{\rm rot}}=\ln({Z_{\rm rotating} / Z_{\rm non-rotating})}$.  A value greater than zero here favors the rotating model. The significance of the preference for each model (radially varying vs flat; rotating vs non-rotating) is based on the magnitude of $\ln{B_{\rm rad}}$ ($\ln{B_{\rm rot}}$) on Jeffery's scale: (0-1), (1-3), (3-5), (5+), corresponds to inconclusive, weak, moderate, and strong evidence in favor of the radially varying (rotating) model. Likewise, the corresponding negative values offer varying degrees of evidence in favor of the flat (non-rotating) model. $\ln{B_{\rm rad}}$ and $\ln{B_{\rm rot}}$ for each galaxy in this work can be found in columns 10 and 11 of Table \ref{tab:tableobs}. For all other parameters estimated by our model, we list the parameter corresponding to the preferred model (flat vs radially varying). Note that only two galaxies (Aquarius and NGC 6822) prefer the flat rotation model.

We compute the posterior distribution with a Multi-Nested Sampling routine \citep{Feroz2008, Feroz2009}. This method directly calculates the evidence and, as a by-product, samples the posterior distribution (for a review of Bayesian method and model comparison see \citealt{Trotta2008}). We marginalize over the prior ranges: $-20 < \overline{v} - v_\mathrm{g} < +20~\kms$, $0 < \sigma < +75~\kms$, $-50 < v_o < +50~\kms$, $0 < \theta < +\pi$, 
$-300 < \mu_{\alpha}  - \bar{\mu}_{\alpha,{\rm HST}} < +300~{\rm mas \, century}^{-1}$, and $-300 < \mu_{\delta}   - \bar{\mu}_{\delta,{\rm HST}} < +300~{\rm mas \, century}^{-1}$, where $v_\mathrm{g}$, $\bar{\mu}_{\alpha,{\rm HST}}$, and $\bar{\mu}_{\delta,{\rm HST}}$ are the values for each galaxy taken from the literature. For the radially varying model, $-1 < \log_{10}{\left(R_o/{\rm kpc}\right)} < \log_{10}{\left( 1.5\times R_{{\rm spectra \, max}} \right)}$.
For several galaxies, we examine larger ranges of $\overline{v}$, $\mu_{\alpha}$, and $\mu_{\alpha}$. This is significant only for Sagittarius, where its close position causes its best fit HST proper motions to be well outside the range considered for other dwarfs. For galaxies with rotation axes near 0 or $\pi$, we marginalize over $-\pi/2 < \theta < +\pi/2$.
All priors are uniform except $\mu_{\alpha}$ and $\mu_{\delta}$, which are Gaussian and centered on the HST measurements. We test our method with mock data sets and verify that the input parameters are recovered. 

Properties taken from the literature and parameter estimates for each observed galaxy in our analysis, given observational dataset $D(\rm v_j, \epsilon_j, \theta_j)$ for each star, can be found in Table \ref{tab:tableobs}. Before moving on to our broad results (Section 5) we will first comment on several galaxies of particular interest in comparison to past work in the literature. 

\subsection{Comments on Individual Galaxies}
\label{sec:indi}

{\bf Draco:} Two recent proper motion measurements for Draco differ by several standard deviations: ($17.7 \pm 6.3,-22.1\pm 6.3$) \citep{Pryor2015} and ($-28.4 ± 4.7, -28.9 \pm 4.1$) \citep{Casetti-Dinescu2016}. \citet{Casetti-Dinescu2016} discuss possible reasons for the discrepancy, but are unable to determine one. We run our analysis with both measurements, which lead to $\vrot$ values of $2.62_{0.90}^{+0.86}$, and $\ratio=0.29_{-0.10}^{+0.10}$ respectively. The two measurements of $\ratio$ are within one sigma of one another and have roughly the same kinematic position angle ($101\degree$ versus $105\degree$). We use the HST measurement to be consistent with the reminder of the classical dSph.

{\bf Sagittarius:} \citet{Penarrubia2010} predict significant rotation in this galaxy based on simulations aimed at reproducing the Sagittarius stream. However, they assumed its progenitor was a late-type disk galaxy ($\vrot\approx20~\kms$). Follow-up work by \citet{Penarrubia2011} did not detect rotation of this magnitude and could only reproduce the line-of-sight velocities observed today using progenitor models with no or little rotation. Similar searches for rotation in Sagittarius have made no conclusive detection \citep{Ibata1997, Frinchaboy2012}.

Our result show very strong evidence for some rotation ($\ratio \simeq 0.28_{-1.33}^{+0.89}$; $\ln{B_{\rm rot}} = 80.26$) but this determination is complicated by the large field of view occupied by Sagittarius on the sky. There are three different proper motion measurements \citep{Dinescu2005, Pryor2010, Massari2013}. All three are discrepant and were obtained from analyzing different fields within Sagittarius. It is possible that the discrepancy is due to the 3D perspective motion or the internal motions of stars within the galaxy. In our analysis, we use the transform of the three measurements into the center of mass frame computed by \citet{Massari2013}:  $-301 \pm 11$, $-145 \pm 11 ~{\rm mas \, century}^{-1}$.

The kinematic axis preferred in our analysis is $\theta=-64\pm6\degree$, which is offset from the photometric major axis of $\theta=102\pm2\degree$ \citep{McConnachie2012}. A velocity gradient along the major axis is expected based on the 3D motion of Sagittarius \citep{Penarrubia2011, Frinchaboy2012}.  It is peculiar, then, that our model favors attributing part of the gradient to rotation instead of the perspective motion. Part of the signal could be induced by tidal interactions, but a more in-depth analysis of the Sagittarius system is required to make a strong conclusion. Another origin of this problem could be the fact that Sagittarius may suffer from a higher degree of foreground contamination from Milky Way stars. We distrust our $\ratio$ analysis for these reasons, and exclude Sagittarius from all figures. However, we note that our estimated value suggests that Sagittarius is not rotationally-supported, and it would lie in the same general region as most of the dSphs analyzed in this work.

{\bf And II:} \citet{Ho2012} detect $\vrot=8.6\pm1.6~\kms$ along the minor axis and a maximum $\vrot=10.9\pm 2.4~\kms$ located at $\theta=113\pm9\degree$ \citep[the photometric position angle is $\theta = 46\pm6\degree, $][]{Ho2012}.  Our kinematic axis is offset from this value:  $\ratio=1.43_{-0.17}^{+0.18}$; $\vrot=11.43_{-1.33}^{+1.31}~\kms$; $\theta=-26\pm4\degree$.  We detect stellar rotation at strong significance near the minor axis, which could have been caused by a minor merger \citep{Amorisco2014}.

{\bf Tucana:}  \citet{Fraternali2009} suggest that a flat rotation curve with $\vrot \approx 15~\kms$ along the major axis is consistent with their data \citep[$\theta=97\degree$,][]{Saviane1996}.  Our analysis finds no evidence for rotation and prefers a value consistent with zero: $\ratio=0.22_{-0.39}^{+0.44}$; $\vrot=4.79_{-8.64}^{+8.99}~\kms$;  $\ln{B_{\rm rot}}=-0.25$.  The position angle is quite unconstrained: $\theta=-6_{-49}^{+59}$.  If Tucana is rotating, a larger sample size will be required to uncover it. 

{\bf Aquarius:} This galaxy has one of the largest preferred $\ratio$ values in our sample ($\simeq 1.70_{-1.01}^{+1.23}$), though the error is large and the Bayesian evidence is weak ($\ln{B_{\rm rot}} =0.62$).  As with Leo A, a larger sample size will be required to make a stronger statement about the rotation and to confirm that it is indeed rotationally-supported.  The kinematic axis of the H{\sc i} gas is at $\theta\approx70\degree$ \citep{Begum2004}.  Our kinematic axis is misaligned at $\theta\approx-1\degree$. The magnitude of the stellar rotation is similar to the observed gas rotation.  

{\bf Leo A:} Although our model prefers a fair amount of rotation in this galaxy ($\ratio =1.99_{-1.09}^{+0.99}$), our analysis yields only weak evidence for rotation in Leo A compared to a non-rotating model ($\ln{B_{\rm rot}} = 1.50$). There is no rotation seen in H{\sc I} gas \citep{Young1996}.  Our potential rotation at  $\theta\approx33\degree$ is almost perpendicular to the H{\sc I} disc at $\theta=102\degree$.  A larger kinematic sample size will be required to make a stronger statement about the rotation.

{\bf Pegasus:}  Stellar rotation in Pegasus was first measured in \citet{Kirby2014} with a magnitude of $\sim 10~\kms$ across the major axis \citep[located at a position angle of 122\degree, ][]{Hunter2006}.  
We measure a larger value that is $20\degree$~offset from the major axis: 
$\vrot=16.25_{-2.24}^{+2.56}$, $\theta=146^{+16}_{-20}$.  A velocity gradient is observed in H{\sc I} across the major axis. It has been suggested that this gradient could be the result of random motions \citep{Young2003}, but since the stellar rotation is detected at such high significance ($\ln{B_{\rm rot}}=29.09$), it seems likely that the gas is rotating as well.  This is in general agreement with the conclusions of \citet{Kirby2014}.

\begin{figure*}
	\centering
	\begin{minipage}{0.475\textwidth}
		\centering
		\includegraphics[scale=0.25]{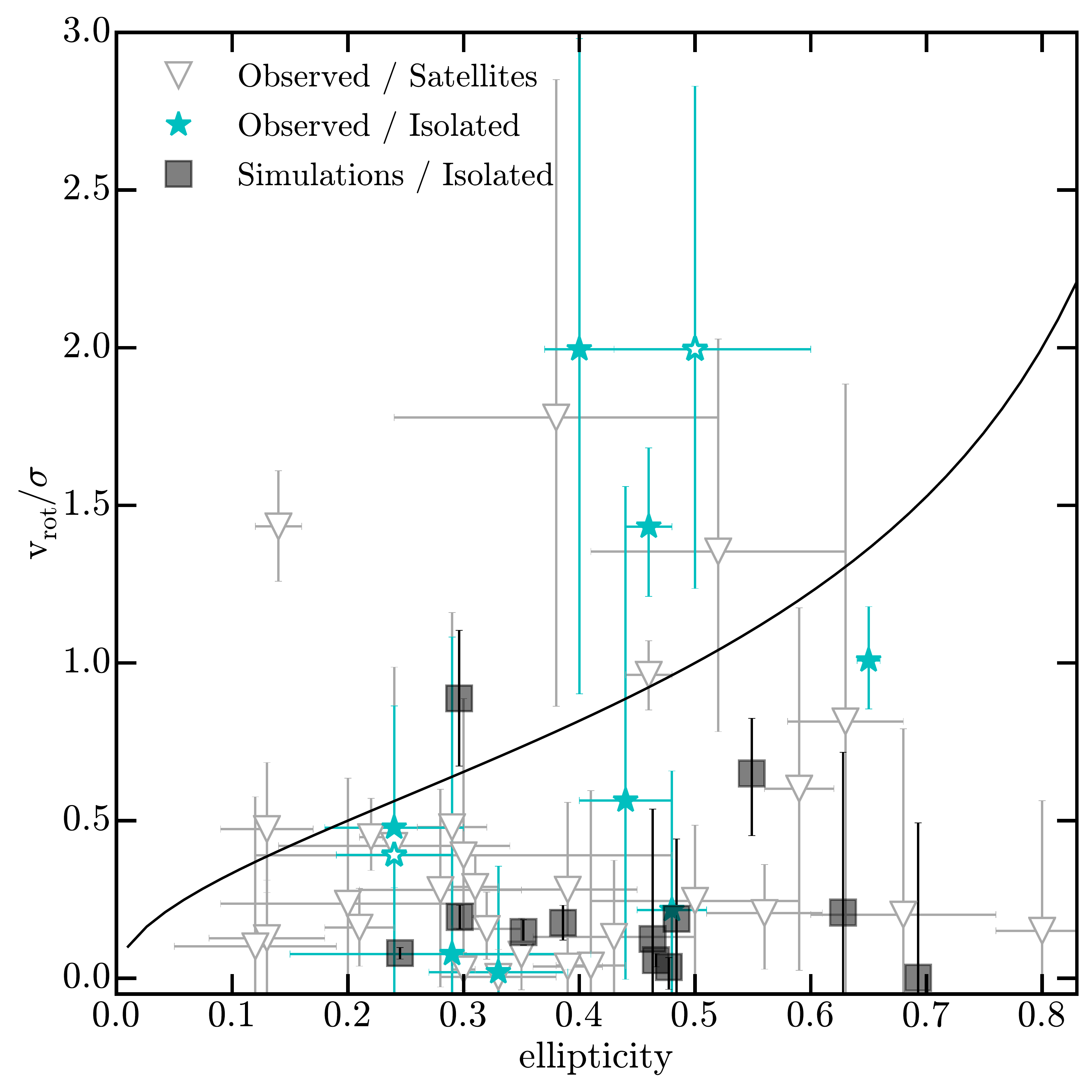}
		\caption{Stellar rotation support $\ratio$ vs. $e$ (ellipticity) for observed satellites of the Milky Way and M31 (open gray triangles), isolated Local Group Dwarfs (cyan stars), and simulated isolated (dIrr) galaxies (gray squares). Open stars show the two galaxies for which a flat rotation model is preferred. The solid line shows the approximate value of $\ratio$ for self-gravitating objects that are flattened by rotation \citep{Binney1978}. The (5/30) observed satellite galaxies (open triangles) that lie above the curve are Andromeda VII, Andromeda II, Coma Berenecis NGC 147, and Canis Venatici II. Only three isolated observed galaxies lie above the curve. Those are Leo A, Pegasus, and Aquarius.} 
		\label{fig:f1}
	\end{minipage} 
	\hfill
	\begin{minipage}{0.475\textwidth}
		\centering
		\includegraphics[scale=0.25]{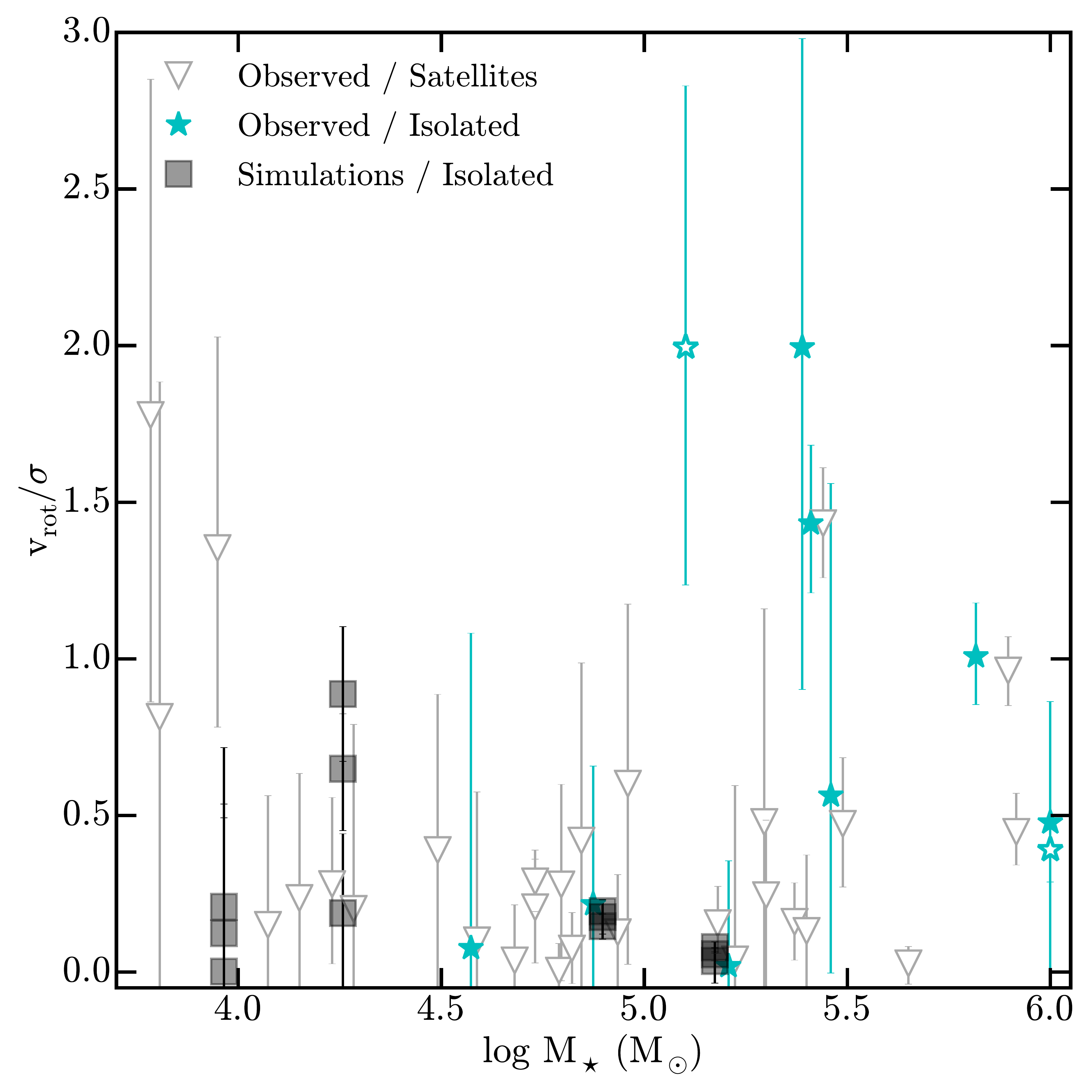}
		\caption{Stellar rotation support ($\ratio$) vs. stellar mass for observed satellites of the Milky Way and M31 (open gray triangles), isolated Local Group Dwarfs (cyan stars), and our simulations (gray squares). No clear trend with stellar mass is seen in the data, although there could be a slight upward turn at the highest masses observed.}
		\label{fig:f2}
	\end{minipage}
\end{figure*}

{\bf WLM:} We measure: $\vrot=14\pm1.6~\kms$, $\sigma=15.6\pm0.9~\kms$, and $\theta=163_{-19}^{23}$.  The position angle we prefer agrees well with the value of $\theta=173\degree$ reported by \citet{Leaman2012}.  In addition, \citet{Leaman2012} measure a velocity dispersion for WLM that is broadly consistent with our value ($\sigma\approx 15~\kms$), and they report a stellar rotation and $\ratio$ that are also consistent with our measured values ($\vrot \sim 15~\kms$; $\ratio = 1.01_{-0.15}^{+0.17}$ with strong evidence).

{\bf IC 1613:} Rotation support in IC 1613 is quite unconstrained in our model, $\ratio=0.48_{-0.63}^{+0.39}$.  Our measured rotation is roughly half as large as that suggested by its H{\sc i} kinematics \citep{Oh2015}. IC 1613 is currently undergoing substantial star formation \citep{Zhang2012, Hunter2012} and has H{\sc i} bubbles and shells \citep{Lozinskaya2002, Silich2006}. The lack of clear rotation support may be due to this starburst phase \citep{Read2016}.

{\bf NGC 6822:} We find moderate evidence for stellar rotation in RGB stars in this galaxy ($\ln{B_{\rm rot}}=3.30$) but the rotation is sub-dominant to the velocity dispersion with $\ratio = 0.41_{-0.15}^{+0.12}$.  The rotation axis is offset from the photometric position angle \citep[located at $\theta=65\degree$, ][]{Battinelli2006} by $\approx 40\degree$:  $\vrot=9.38_{-3.46}^{+2.77}~\kms$, $\theta=108_{-11}^{+9}$, $\sigma=22.62_{-0.92}^{+ 0.99}$. Stellar rotation in Carbon stars was previously detected along the major axis \citep{Demers2006}.  As the  H{\sc I} disk is perpendicular to the stellar component, they label NGC 6822 as a polar ring galaxy.  N-type Carbon stars have variable velocity, limiting the precision of the \citet{Demers2006} measurements to $\pm15~\kms$.  In addition, their sample was created from two telescopes, with a velocity offset of $46~\kms$ between each measurement and $\Delta \overline{v} \approx20$ between the RGB stars and C stars.  With these caveats, it is intriguing that the different tracers all have a different kinematic axes, possibly hinting at past mergers. \citet{Valenzuela2007} model NGC 6822 using a tilted ring analysis and show that the presence of a bar can artificially decrease the rotation signal for some projections. We do not account for bars in our model, instead opting to use the same analysis for each galaxy in our sample.

\subsection{Simulation analysis}

We apply an identical method for calculating $\ratio$ to the simulations (see Section \ref{sec:method}). To calculate the ellipticity values for the simulations, we use a simple method outlined in \citet{Cappellari2007} for converting two dimensional field data to a single $\ratio$ value. For each of the three orthogonal distributions, the galaxy is rotated along the axis parallel to the line-of-sight until there is a maximum in the difference between velocity measurements in the left and right hemispheres of the projection plane. Then, after binning the stars in two dimensions, we sum up the effective ``flux" in each bin and weight the bins by their distance from the center of the simulated galaxy, according to this formula:
\begin{equation}
(1-e)^2 = \frac{\sum_{n=1}^{N} F_n y_n}{\sum_{n=1}^{N} F_n x_n},
\end{equation}
where $x_n$ and $y_n$ are the bin centers and we replace flux, $F_n$, with the number of star particles in that bin. \footnote{We have tested that this method produces ellipticity values consistent with those obtained by performing a 2D Gaussian fit to histograms of the ``flux" (in this case the number of star particles) in a $2\times 2$ grid along the line-of-sight to each object.} All analysis on the simulations is done on all star particles within $3~\kpc$ of the center of each simulated galaxy. This choice allow us to select all stars that belong to the main galaxy while excluding any satellites.

\section{Results}
\label{sec:res}

Figure \ref{fig:f1} shows $\ratio$ vs. $e$ (ellipticity) for all objects in our study. $\ratio$ is a standard diagnostic for detecting rotational support in more massive systems \citep{Bender1993} as well.
Observed Milky Way and M31 satellites are shown as open triangles, observed isolated dwarfs are shown as cyan stars, and simulated (isolated) galaxies are gray rectangles. The black line shows the expectation for self-gravitating objects flattened by rotation \citep{Binney1978}. For the sake of concreteness, we consider objects that lie above this line to be at least marginally rotationally-supported. The galaxy ellipticity values were drawn from the literature.

Of all the galaxies in our sample, eight have $\ratio$ values that are consistent with being supported by rotation, rather than dispersion: Coma Berenecis, Canis Venatici II, Andromeda II, Andromeda VII, NGC 147, Aquarius, Leo A and Pegasus. Of these, only And II (dSph), NGC 147 and Pegasus (dIrr) show rotation at strong significance. The Bayesian evidence that Aquarius and Leo A are rotating is inconclusive or weak -- the small sample sizes prohibit a stronger statement. We also detect sub-dominant rotation at strong significance in NGC 185, Sagittarius, and WLM.  We detect some (sub-dominant) rotation in NGC 6822, but at a lower significance.

\begin{figure*}
	\centering
	\includegraphics[scale=0.39]{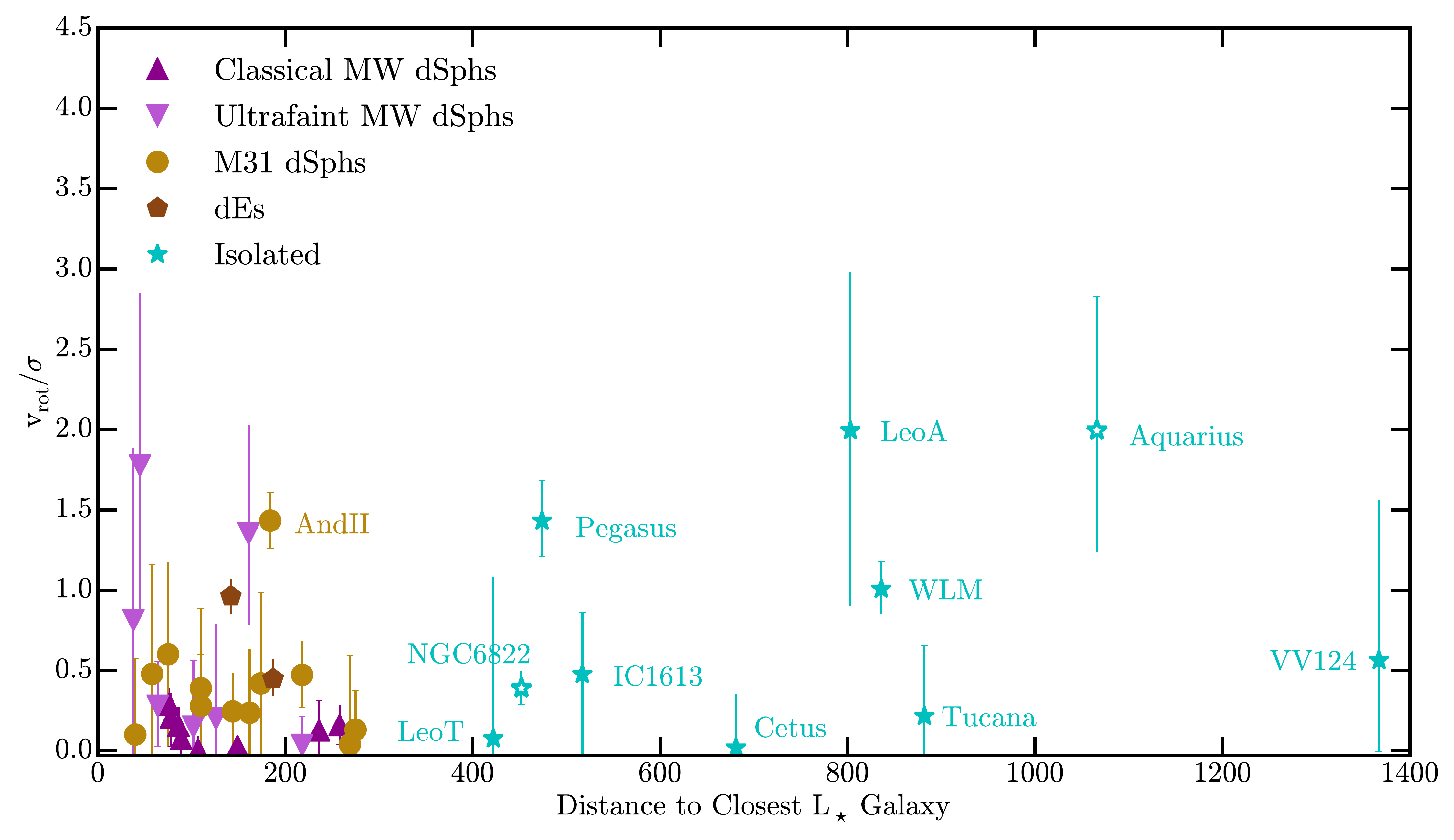}
	\caption{Rotation support $\ratio$ vs $\rm d_{\rm L_\star}$, distance from the dwarf to the closest $L_{\star}$ galaxy (either the Milky Way or M31), for observed classical Milky Way dSphs (up-facing dark magenta triangles),  ultra-faint dSphs (down-facing light magenta triangles), M31 dSphs (gold circles), isolated Local Group Dwarfs (cyan stars) and dEs (brown pentagons). There is no clear trend between $\ratio$ and $\rm d_{\rm L_\star}$, as predicted by tidal stirring models.
	}
	\label{fig:dist}
\end{figure*}

Perhaps the most striking feature of Figure \ref{fig:f1} is the distribution of isolated galaxies. 7/10 of the isolated dwarfs in our analysis have $\ratio$ vs ellipticity values that are consistent with being dispersion-dominated, while (6/10) have $\ratio \lesssim 1.0$. All have $\ratio \lesssim 2.0$ This is in stark contrast to the common assumption that dIrrs have stellar disks that are smaller versions of their more massive, rotating counterparts. Even the three rotation-dominated systems are only modestly so, with $\ratio \simeq  1.5-2$,  which is significantly less rotation than a canonical cold disk, and below the values typically assumed as initial conditions for tidal stirring scenarios for dSph formation ($\ratio \gtrsim 3$) \citep{Kazantzidis2011a}.

Our simulated dwarfs are shown as filled gray squares, each displayed at three orthogonal (but random) orientations (for a total of 12 points).  The range of simulated ellipticities is consistent with the range of the observed dwarfs. Our simulated dwarfs also have $\ratio$ values that are broadly consistent with the data~\footnote{The simulations also show a higher degree of rotation in their cold gas, in qualitative agreement with observations \citep{Mateo1998, Grebel1999}.}.  We will need more simulations \citep{Fitts2016} to determine whether we can ever achieve the modest fraction of isolated galaxies (3/10) with $\ratio \simeq 1.5-2$ that we see among isolated Local Group dwarfs. If not, then this may suggest that the star formation is too bursty, or that the specific feedback implementation causes too much coupling between the injected energy and both the stellar populations and the dissipationless dark matter at the hearts of dwarf galaxies \citep{Onorbe2015, Chan2015}.   

Figure \ref{fig:f2} shows $\ratio$ vs. stellar mass for all of the objects in our sample. No obvious trend with stellar mass is seen, though we note that 4/6 systems with $\ratio > 1$ all have $M_\star > 10^6 M_\odot$. \citet{Kormendy2009} show that more luminous ($ -23.24  < M_{\rm V} < -15.53$) dSphs in the Virgo cluster form an extension of Local Group dSphs in the Sersic index-$M_{\rm V}$ plane, and \citet{Toloba2014} find a wide range of $\ratio$ values for subset of the Virgo dwarfs ($-19.0 < M_{\rm r} < -16.0$), but both of these studies rely on photometry from diffuse light. Extending our analysis (using resolved stellar populations) to higher mass objects, both observed and simulated, would be useful in detecting either a trend between of $\ratio$ and $\mstar$ at higher mass, or a discontinuity between dSphs/dEs and rotating disks. However, at least on the observational side, this analysis may have to wait for the next generation of telescopes. An initial analysis of one slightly more massive ($\mstar \sim 10^9~\msun$) simulated dwarf run with the same code at slightly lower resolution, shows that it is also dispersion-supported ($\ratio \lesssim 0.25$), but more runs at higher mass are needed in the simulations to make a stronger statement about mass trends.

\subsection{Inferring 3D Rotation}
\label{sec:threeD}

The $\ratio$ values estimated by our model and listed in Table \ref{tab:tableobs} are lower limits to the intrinsic amount of rotation support for each galaxy. This is due to the fact that the line of sight velocity we measure is  $\vrot = v^{{\rm intrinsic}}_{{\rm rot}} \sin{i}$. We can correct for the actual measured inclination of at least those (6/10) galaxies with measured inclination angles in H{\sc I}: Aquarius ($66.7\degree$), Pegasus ($69.4\degree$), WLM ($74.0\degree$), IC1613 ($48.0\degree$)\citep{Oh2015}, Leo A ($60.3\degree$) \citep{Hunter2012}, and NGC 6822 ($60\degree$) \citep{Weldrake2003}. Additionally, \citet{Read2016} estimate an inclination of $20\degree$ for Leo T by matching their simulations to the galaxy's photometric light profile. With these inclination values, the estimated $\ratio$ value for Aquarius changes from $1.99$ to $2.17$, for IC 1613 from $0.48$ to $0.64$, for Leo A from $1.99$ to $2.30$, for NGC 6822 from $0.39$ to $0.45$, for Pegasus from $1.43$ to $1.53$, for WLM from $1.01$ to $1.05$, and for Leo T from $0.08$ to $0.22$. None of the estimated $\ratio$ values move from $<1$ to $ > 1$ and, of the four galaxies without measured inclinations, VV 124 has the highest $\ratio$ value ($0.56$), and would still have $/ratio < 1$ for inclination as low as $35\degree$. Therefore it is unlikely that inclination severely affects the primary result presented in this work -- that dwarf galaxies do not form as cold, rotating disks with $\ratio \gtrsim 2$.

\begin{figure}
	\includegraphics[scale=0.25]{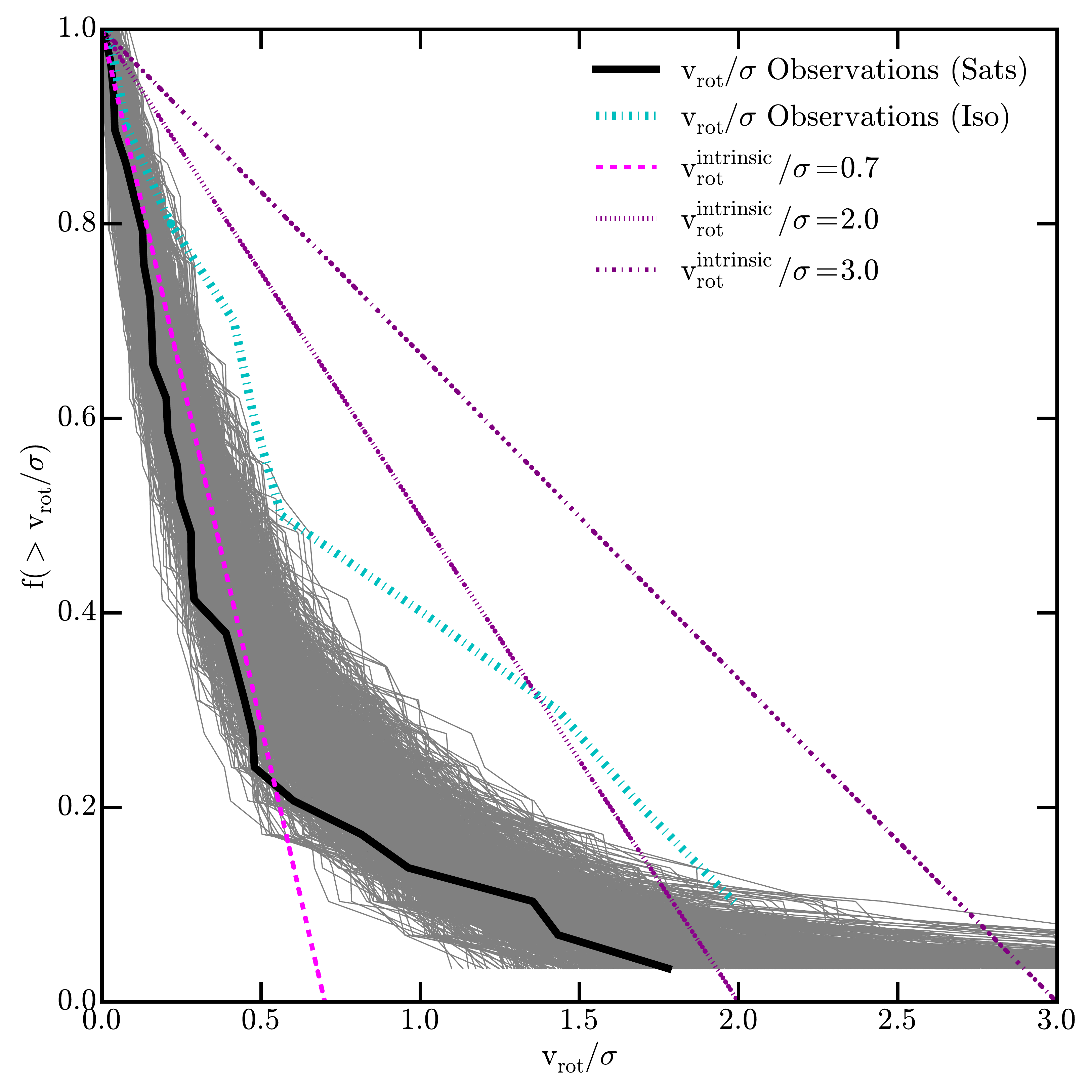}
	\caption{Distribution of measured $\ratio$ on the sky if all galaxies are assumed to have an intrinsic $\ratio$ value of $0.7$ (dashed magenta line), $2.0$ (dotted magenta line) or 3.0 (thin dash-dotted magenta), but are viewed with a random inclination. The thick solid black (thick dash-dotted cyan) line shows the distribution of estimated median $\ratio$ values for the satellites (isolated dwarfs) in our sample. The error in the satellite distribution is illustrated with 1000 thin gray lines, each consisting of points drawn randomly from Gaussian distributions with parameters taken from the estimated values for each of the 29 dSphs in our sample.  For clarity, the error is not shown for the distribution of isolated galaxies, but has a wider spread than the lines for the satellite population. The isolated distribution is distinct from the satellite distribution, and more closely tracks $v^{{\rm intrinsic}}_{{\rm rot}} / \sigma \sim 1-2$. However, it is clear that the $v^{{\rm intrinsic}}_{{\rm rot}} / \sigma$ values for the isolated galaxies are much less than 3, the commonly used value in tidal stirring simulations.
	}
	\label{fig:incl}
\end{figure}

Another way of evaluating the effect of inclination on the estimated line of sight $\ratio$ values is to infer something about the distribution of
three dimensional rotation in our sample by comparing the observed distribution ($\ratio = v^{{\rm intrinsic}}_{{\rm rot}} \sin{i} / \sigma$) to what would be measured for a given $v^{{\rm intrinsic}}_{{\rm rot}} $ viewed in projection from random orientations. \footnote{We have assumed that $\sigma$ is independent of viewing angle, which is a good approximation for dispersion-supported objects. For rotationally supported objects, if $\sigma$ is larger in the plane of the disk compared to vertically, as is the case for the Milky Way, the result of a face-on viewing angle will have less of an effect than described here.} As can be seen in Figure \ref{fig:incl}, the observed distribution of $\ratio$ for satellite galaxies (thick solid black line) closely matches the distribution of $v^{{\rm intrinsic}}_{{\rm rot}} / \sigma = 0.7$ (dashed magenta line), but with a slight tail out to higher intrinsic rotation values. The multiple (1000) gray lines indicate the possible distributions of $\ratio$ if each data point is selected from a Gaussian distribution centered on the $\ratio$ values from our model, and with standard deviations also taken from the model ($1 \sigma$ errors).\footnote{For clarity, the error is not shown for the distribution of isolated galaxies, but has a wider spread than the error in the satellite population.} The distribution of line of sight $\ratio$ values for the isolated dwarfs (thick dash-dotted cyan line) lies just outside of this ``error band" for satellite galaxies, with a distinct excess at $\ratio \sim 0.7 - 2.0$. The isolated galaxy distribution more closely matches a distribution of $v^{{\rm intrinsic}}_{{\rm rot}}/ \sigma \approx 1-2$ (dotted magenta line at $v^{{\rm intrinsic}}_{{\rm rot}}/ \sigma = 2$), but falls far short of matching the (thin dash-dotted magenta) $v^{{\rm intrinsic}}_{{\rm rot}}/ \sigma =3$ line. While it appears that the isolated sample has more intrinsic rotation than the satellite sample, the isolated sample remains only marginally rotationally supported, with none as cold as $v^{{\rm intrinsic}}_{{\rm rot}} / \sigma \sim 3$, the value commonly used in tidal stirring simulations.

\section{Discussion}
\label{sec:dis}

A clear prediction made by the tidal stirring model of dSph formation is the increase of $\ratio$ with increasing distance from a more massive galaxy \citep{Kazantzidis2011a}. Because the most distant galaxies in the Local Group could have had no more than one pericenter passage in a Hubble time \citep[and most are expected to have had none, e.g.,][]{Garrison-Kimmel2014a}, we would expect that galaxies that lie beyond the virial radius of either giant to have larger $\ratio$ values if tidal stirring plays the primary role in shaping dwarf galaxy dynamics.

Figure \ref{fig:dist} explores this possibility by showing $\ratio$ vs. distance from the closest massive Local Group galaxy (MW or M31). We do not see any clear trend between $\ratio$ and distance to a massive galaxy, as would be expected if multiple close pericenter passages were necessary for removing rotation from dwarf galaxies.   

Given the lack of trend between $\ratio$ and distance, we are more inclined to suspect that the stars in small galaxies are formed in a medium with marginal rotation support, and undergo merely a modest transformation to become dSphs. Some further evidence for this comes from \citet{Sanchez_Janssen2010}, who study $11,753$ galaxies from the Sloan Digital Sky Survey (SDSS) and \citet{Karachentsev2004}. They suggest the existence of a critical stellar mass, $\mstar = 2 \times 10^9~\msun$, below which all galaxies become systematically thicker. One important question that will need to be investigated with future simulations is whether or not galaxies that start out with $\ratio \sim 1-2$ can undergo enough of a transformation to match the near zero values observed for the smallest dwarf satellites within the infall time constraints provided by cosmological simulations. Although it is likely that a mild transformation in $\ratio$ would take much less than the $10~\gyr$ required by tidal-stirring simulations, it would be instructive to use mildly dispersion-dominated dwarfs -- in particular at slightly higher mass than those presented here -- as the initial conditions for those models. An initial study has been performed by \citet{Mayer2011}, who perform tidal stirring simulations on a gas-dominated, cosmological dwarf with a larger vertical scale height (aspect ratio $\sim 3:1$) resulting from stellar feedback-driven turbulence in the star-forming gas. In this work, the thicker dwarf reaches $\ratio < 0.5$ in just under 2 pericenter passages, without the typical bar formation and subsequent buckling common in tidally-induced transformations. Additional work along these lines should prove particularly informative.

We have checked to see if the four observed ``rotating" systems are distinct in other properties that might help explain why they have $\ratio$ values that are $ > 1$. These objects do not appear to be significant outliers in metallicity, inner density, star formation history or star formation rate, but a more thorough search for galaxy properties that do correlate with $\ratio$ would be useful. In addition to explaining the small number of outlying observed dwarfs, it could further explain why most of the simulated galaxies fail to demonstrate an elevated $\ratio$ -- perhaps all simulated halos were selected in a way that disfavors the property that best correlates with rotation support. \citet{Gallart2015} do find that Aquarius and Leo A qualify as ``slow" dwarfs, having formed in a low density environment which leads to a small fraction of their stars forming early, followed by continues star formation until the present time. This is in contrast to ``fast" dwarfs that form the majority of their stars in a single, early burst. However, our simulated dwarfs were preferentially selected to inhabit low density environments, and yet have low $\ratio$ values. Additional simulations selected from a variety of environments would be useful to test these effects.

All stars analyzed in this work are either red giant or horizontal branch stars, so it is unlikely that we are biasing our sample due to stellar ages. A separate analysis of stellar populations with varying ages -- in both the observations as well as the simulations -- would likely be informative, but is beyond the scope of this paper.

\section{Summary and Conclusion}
\label{sec:summ}

We have performed a systematic Bayesian search for stellar rotation in 40 dwarf galaxies ($10^{3.5} \msun < \mstar < 10^{8} \msun$) in the Local Group, using resolved stellar kinematic data from the literature.  We find that the vast majority of these galaxies ($\sim 80\%$) have $\ratio$ values that imply
dispersion-supported kinematics.   In particular, we find that 6/10 isolated dwarfs in our sample have
$\ratio < 1.0$, and all have $\ratio \lesssim 2$ (see Figure 1 and Table 1).  This result for the most distant LG dwarfs galaxies contrasts the common assumption that dwarf galaxies form with cold, rotationally-supported stellar disks (with $\ratio \sim 3$).  We find no strong trend of $\ratio$ with $\mstar$ within the mass range studied (Figure 2), nor any trend of $\ratio$ with distance from large host galaxy in the Local Group (Figure 3), as would be expected if tidal stirring scenarios drive a kinematic transformation of stars in dIrr galaxies to dSph galaxies over multiple pericenter passages.

Taken together, our results suggest that dwarf galaxies form as puffy stellar systems that either dispersion-supported, or only mildly rotation-dominated. The conversion of a dIrr galaxy into dSph galaxy may involve little more than the removal of its gas, and a resulting mild decrease of its $\ratio$. Specifically, the process of gas stripping itself may be enough to shock the potential, transforming a stellar system with $\ratio \sim 1.5$ into a system with $\ratio \sim 0.5$. Detailed simulations of this kind will be needed to test this hypothesis.

The formation of initially dispersion-supported systems is more likely to occur within dark matter halos with shallow potential wells (KWB), especially if explosive feedback effects act to dynamically heat stellar populations after the stars form. We have examined $\ratio$ in four cosmological zoom-in simulations of isolated dwarf galaxies that include such explosive feedback events \citep{Wheeler2015,Onorbe2015, Muratov2015}. These simulated dwarfs have $\mstar - \mhalo$ values that lie very close to extrapolated abundance-matching relations \citep{Hopkins2014a, Onorbe2015, Wheeler2015}, so the total amount of energy injected to the surrounding medium is likely appropriate. However, the strength and frequency of bursts could modify the fraction of energy that couples to stars and dark matter, and so could be driving the stellar kinematics. All but 2 of the 12 viewing angles for the simulated dwarfs show (mock-observed) stellar dispersion support values $\ratio \simeq 0 - 0.8$ (and ellipticities $\simeq 0.2 - 0.7$), and all are completely consistent with our derived properties of observed satellite dwarfs and isolated dwarfs without a significant need for harassment from a massive neighbor. While these simulations are certainly not the final word on the formation of dwarf galaxies, the result suggests that it is at least reasonable to posit that dwarf galaxies are generally born moderately hot and are never strongly rotationally-supported.

The comparison between our model isolated dwarfs and the data did reveal one source of potential tension: none of our simulated dwarfs have stellar rotations that are as high as the highest in our sample (the 3/10 isolated galaxies with $\ratio \simeq 1.5 - 2$). This is not particularly surprising, given the small number of simulations analyzed here, but if this discrepancy holds in the face of better data and more simulations, it could point to a new test for feedback models. In particular, it is via bursty and violent feedback episodes that the dark matter cores in these halos are reduced in density, thus alleviating potential problems with \lcdm~like  the Too Big to Fail problem \citep{Boylan-Kolchin2011}. As first pointed out by \citet{Teyssier2013}, the same outbursts also inject significant random energy into the stellar populations \citep[see also ][]{Kawata2014, Chan2015}.  A more detailed comparison between simulated and observed $\ratio$ values may offer an interesting direction in testing models that attempt to solve dark matter problems via explosive feedback episodes \citep[e.g.][]{Governato2012G,Teyssier2013,Brooks2014,Onorbe2015,Chan2015}. Can these same models preserve the mild stellar rotation that is seen in some isolated dwarfs? Or, is stellar rotation only seen in galaxies with cuspy density distributions, which would be an important prediction of such models? The analysis of observational data provided here will hopefully provide an important benchmark for this question going forward.

\section*{Acknowledgments} 
We thank the anonymous referee, for comments that made the paper clearer and more robust. We also thank Josh Simon, Marla Geha, Ngoc Nhung Ho, Nicolas Martin, Serge Demers, Evan Kirby and Ryan Leaman for generously providing their data for this project. We further thank Ryan Leaman and Evan Kirby for very helpful discussions.
This work used computational resources granted by NASA Advanced Supercomputing (NAS) Division, NASA Center for Climate Simulation, Teragrid and by the Extreme Science and Engineering Discovery Environment (XSEDE), which is supported by National Science Foundation grant number OCI-1053575 and ACI-1053575, the latter through allocation AST140080 (PI: Boylan-Kolchin). CW acknowledges support from the Josephine de Karman Fellowship Trust. CW and JB were supported in part by \textit{Hubble Space Telescope} grants HST-AR-13921.002-A and HST-AR-13888.003-A. MBK acknowledges support from NASA through \textit{Hubble Space Telescope} theory grants (programs AR-12836 and AR-13888) from the Space Telescope Science Institute (STScI), which is operated by the Association of Universities for Research in Astronomy (AURA), Inc., under NASA contract NAS5-26555. MBK and AF acknowledge support from NSF grant AST-1517226. OE acknowledges funds from NSF grant NSF-AST 1518291 and \textit{Hubble Space Telescope} grant HST-GO-13343.09-A. Support for PFH was provided by an Alfred P. Sloan Research Fellowship, NASA ATP Grant NNX14AH35G, and NSF Collaborative Research Grant \#1411920 and CAREER grant \#1455342. Some numerical calculations were run on the Caltech compute cluster ``Zwicky'' (NSF MRI award \#PHY-0960291) and allocation TG-AST130039 granted by the Extreme Science and Engineering Discovery Environment (XSEDE) supported by the NSF. DK received support from XSEDE allocation TG-AST-120025 (PI: Kere\v{s}) National Science Foundation grant number AST-1412153, and funds from the University of California, San Diego.


\label{lastpage}
\end{document}